\documentclass{pnastwo}

\usepackage{graphicx,psfrag,xspace}

\usepackage{amssymb,amsfonts,amsmath}

\contributor{Submitted to Proceedings of the National Academy of Sciences of the United States of America}
\url{www.pnas.org/cgi/doi/10.1073/pnas.0709640104}
\copyrightyear{2008}
\issuedate{Issue Date}
\volume{Volume}
\issuenumber{Issue Number}

\begin{document}



\title{Spatial extent of an outbreak in animal epidemics}





\author{Eric Dumonteil\affil{1}{CEA/Saclay, DEN/DM2S/SERMA/LTSD, 91191 Gif-sur-Yvette Cedex, France},
Satya N. Majumdar\affil{2}{CNRS - Universit\'e Paris-Sud, LPTMS, UMR8626, 91405 Orsay Cedex, France},
Alberto Rosso\affil{2}{CNRS - Universit\'e Paris-Sud, LPTMS, UMR8626, 91405 Orsay Cedex, France},\and
Andrea Zoia\affil{1}{CEA/Saclay, DEN/DM2S/SERMA/LTSD, 91191 Gif-sur-Yvette Cedex, France},
}

\contributor{Submitted to Proceedings of the National Academy of Sciences of the United States of America}

\maketitle

\begin{article}

\begin{abstract} Characterizing the spatial extent of epidemics at the outbreak stage is key to controlling the evolution of the disease. At the outbreak, the number of infected individuals is typically small, so that fluctuations around their average are important: then, it is commonly assumed that the susceptible-infected-recovered (SIR) mechanism can be described by a stochastic birth-death process of Galton-Watson type. The displacements of the infected individuals can be modelled by resorting to Brownian motion, which is applicable when long-range movements and complex network interactions can be safely neglected, as in case of animal epidemics. In this context, the spatial extent of an epidemic can be assessed by computing the convex hull enclosing the infected individuals at a given time. We derive the exact evolution equations for the mean perimeter and the mean area of the convex hull, and compare them
with Monte Carlo simulations.
 \end{abstract}






\dropcap{M}odels of epidemics traditionally consider three classes of populations, namely, 
the susceptibles (S), the infected (I), and the recovered (R). This provides the basis of the 
so-called SIR model~\cite{bailey, mckendrik}, a {\em fully connected
mean-field} model where the population sizes of the three species evolve with time $t$ via
the coupled nonlinear equations: $dS/dt=-\beta I S$; $dI/dt=\beta I S- \gamma I$ and
$dR/dt= \gamma I$. Here $\gamma$ is the rate at which an infected individual recovers
and $\beta$ denotes the rate at which it transmits the disease to a susceptible~\cite{murray, bartlett, andersson}. In the simplest version of these 
models, the recovered can not be infected again. These rate equations conserve
the total population size $I(t)+S(t)+R(t)=N$,
and one assumes that initially there is only one infected individual and the
rest of the population is susceptible: $I(0)=1$, $S(0)=N-1$, and $R(0)=0$.
Of particular interest is the outbreak stage, i.e., the
early times of the epidemic process,
when the   
susceptible population is much larger than the number of infected or recovered. During this regime, for large $N$, the
susceptible population hardly evolves and stays $S(t)\approx N$, so that nonlinear effects can be safely neglected and one can
just monitor the evolution of the {\em infected population alone}: $dI/dt\approx (\beta N-\gamma)I(t)$. Thus, the ultimate fate of the epidemics 
depends on  
the key dimensionless parameter $R_0=\beta N/\gamma$, which is called the
reproduction rate. If $R_0>1$ the epidemic explodes and invades a finite fraction of the population,
if $R_0<1$ the epidemic goes to extinction,
and in the critical case $R_0=1$ the infected population remains constant~\cite{antal, may, antia}.

This basic deterministic SIR has been generalized to a variety of both deterministic as well
as stochastic models, whose 
distinct advantages and shortcomings are discussed at length in~\cite{whittle, kendall,  
bartlett_book}.
Generally speaking, stochastic models are more suitable in presence of a 
small number of infected individuals, when fluctuations around the average may be 
relevant~\cite{whittle, kendall}.
During the outbreak of epidemics, the infected population is typically small: in this regime, the evolution
can be modeled by resorting to a stochastic birth-death branching process of 
the Galton-Watson type for the number of infected~\cite{whittle, kendall, bartlett_book}, where
each infected individual transmits the disease to another individual at rate $\beta N$
and recovers at rate $\gamma$. The epidemic may become endemic for $R_0>1$, becomes extinct for $R_0<1$, whereas
for $R_0=1$ fluctuations are typically long lived and completely control the
time evolution of the infected population~\cite{murray, bartlett, andersson}. 

How far in space can an epidemic spread? Branching processes alone are not sufficient to 
describe an 
outbreak, and spatial effects must necessarily be 
considered~\cite{bailey, bartlett, spatial_epidemics, radcliffe, wang}. Quantifying the 
geographical spread of an epidemic is closely related to the modelling of the population 
displacements. Brownian motion is often considered as a paradigm for describing the migration of individuals, despite some well-known shortcomings:
for instance, finite speed effects and preferential displacements
are neglected. Most importantly, a number of recent 
studies have clearly shown that individuals geographically far apart can actually be closely related to each 
other through the so-called small-world connections, such as air traffic, public 
transportation and so on: then, the spread of epidemics among humans can not be 
realistically modelled without considering these complex networks of interconnections~\cite{patterns, influenza, colizza, brockmann}. Nonetheless, Brownian motion provides a reasonable basis for studying disease propagation in 
animals and possibly in plants (here, pathogen vectors are 
insects)~\cite{murray}.

While theoretical models based on branching Brownian motion have 
provided important insights on how the population size grows and fluctuates with 
time in a given domain~\cite{bailey, bartlett, radcliffe, wang}, another fundamental question is 
{\em how the spatial extension of the infected population evolves with time}. Assessing the 
geographical area travelled by a disease is key to the control of epidemics, and this is 
especially true at the outbreak, when confinement and vaccination 
could be most effective. One major challenge in this very practical field of disease control is
how to quantify the area that
needs to be quarantined during the outbreak. For animal epidemics, this issue has been investigated 
experimentally, for instance in the case of equine influenza~\cite{equine}. The most popular and widely used
method for this consists in recording the set of positions of the infected animals 
and, at each time instant, construct a {\em convex hull, i.e., a minimum convex polygon}
surrounding the positions (Fig.~\ref{fig1}; for a precise  
definition of the convex hull, see below). 
The convex hull at time $t$ then  
provides a rough measure of the area over which the infections have spread
up to time $t$. The convex hull method is
also used to estimate the home range of animals, i.e., the
territory explored by a herd of animals 
during their daily search for food~\cite{Worton, Luca}.

In this paper, we model the outbreak of an epidemic
as a Galton-Watson branching process
in presence of Brownian spatial diffusion. Despite infection dynamics being
relatively simple, the corresponding convex hull is a rather complex function 
of the trajectories of the infected individuals up to time $t$, whose
statistical properties seem to be a formidable problem.
Our main goal is to characterize
the time evolution of the convex hull associated to this process, in particular its mean perimeter and area.

The rest of the paper is organized as follows. We first describe precisely the model
and summarize our main results. Then, we provide a derivation of our analytical findings, 
supported by extensive numerical simulations. We conclude with perspectives
and discussions. Some details 
of the computations are
relegated to the Supplementary Material.

\section{The model and the main results}

Consider a population of $N$ individuals, uniformly distributed in a two dimensional
plane, with a single infected at the origin at the initial time. At the outbreak, it is sufficient to keep track of the positions of the infected, which will 
be marked as `particles'. The dynamics of the infected individuals 
is governed by the following
stochastic rules. In a small time interval $dt$, each infected alternatively

\vskip 0.2cm

\noindent {\bf (i)} recovers with probability $\gamma\, dt$. 
This corresponds to the death of a particle with rate $\gamma$.

\vskip 0.2cm

\noindent {\bf (ii)} infects, {\em via 
local contact}, a 
new 
susceptible
individual from the background with probability $b\, dt $. This corresponds to the birth of 
a new particle that can subsequently diffuse.
The originally infected particle still remains infected, which
means that the trajectory of the originally infected particle branches into two
new trajectories. The rate $b$ replaces the rate $\beta N$ in the SIR or the
Galton-Watson process mentioned before.

\vskip 0.2cm

\noindent {\bf (iii)} diffuses with diffusion constant $D$ with probability $1- (\gamma + b)\, dt$. The coordinates $\{x(t),y(t)\}$ of 
the particle get updated to the new values $\{x(t)+\eta_x(t)\, dt, y(t)+\eta_y(t)\, dt\}$,
where $\eta_x(t)$ and $\eta_y(t)$ are independent Gaussian white noises with zero mean
and correlators $\langle \eta_x(t)\eta_x(t')\rangle = 2D \delta(t-t')$, 
$\langle \eta_y(t)\eta_y(t')\rangle = 2D \delta(t-t')$ and 
$\langle \eta_x(t)\eta_y(t')\rangle = 0$. 

\vskip 0.2cm

The only dimensionless parameter in the model is the ratio $R_0=b/\gamma$, 
i.e, the basic reproduction number.

Consider now a particular history
of the assembly of the trajectories of all the infected
individuals up to time $t$, starting from a single infected initially at the origin (see Fig.~\ref{fig1}).
For every realization of the process, we construct
the associated convex hull $C$.
To visualize the convex hull, imagine stretching a rubber band
so that it includes all the points of the set at time $t$ inside it
and then releasing the rubber band. It shrinks and finally gets stuck
when it touches some points of the set, so that it can not shrink any further.
This final shape is precisely the convex hull associated
to this set.

In this paper, we show that the mean perimeter $\langle L(t) \rangle$ and the mean area $\langle A(t) 
\rangle$ of the convex hull are ruled by two coupled nonlinear partial differential equations that can be solved 
numerically for all $t$ (see Fig.~\ref{fig2}). The asymptotic behavior for large $t$ can be determined
analytically for the critical ($R_0=1$), subcritical ($R_0<1$) and supercritical ($R_0>1$) regimes.
In particular, in the {\em critical} regime the mean 
perimeter {\em saturates} to a finite value as $t\to \infty$, 
while the mean area {\em diverges logarithmically} for large $t$
\begin{eqnarray}
\langle L(t \to \infty)\rangle &=& 2\pi \sqrt{\frac{6 D}{\gamma}}+ {\cal O}(t^{-1/2}) 
\label{mean_perimeter} \\
\langle A(t \to \infty)\rangle &=& \frac{24 \pi D}{5\gamma}\ln t + {\cal O}(1)  .
\label{mean_area}
\end{eqnarray}
This prediction seems rather paradoxical at a first glance. 
How can the perimeter of a polygon be finite while its area is divergent? 
The resolution to this paradox owes its origin precisely to statistical
fluctuations. The results in Eqs.~[\ref{mean_perimeter}] and~[\ref{mean_area}] are true only 
on average.
Of course, for each sample, the convex hull has a finite perimeter and a finite area.
However, as we later show, the probability distributions of these random variables have 
power-law tails at long time limits. For instance, while ${\rm Prob}(L,t\to \infty)\sim 
L^{-3}$
for large $L$ (thus leading to a finite first moment), the area distribution behaves
as ${\rm Prob}(A,t\to \infty)\sim A^{-2}$ for large $A$. Hence the mean area is 
divergent
as $t\to \infty$ (see Fig.~\ref{fig2}).

When $R_0 \neq 1$, the evolution of the epidemic is controlled by a characteristic time
$t^*$, which scales like $t^* \sim |R_0-1|^{-1}$. For times $t < t^*$ the epidemic behaves
as in the critical regime. In the {\em subcritical} regime, for $t>t^*$ the quantities $\langle L(t) \rangle$ and $\langle A(t) \rangle$ rapidly 
saturate and the epidemic goes eventually to extinction. In contrast, in the {\em supercritical} regime (which is the most relevant for virulent epidemics that spread fast), a new time-dependent behavior emerges when $t>t^*$, since there exists a finite probability (namely $1-1/R_0$) that epidemic never
goes to extinction (Fig.~\ref{fig3}). More precisely, we later show that
\begin{eqnarray}
\langle L(t \gg t^*)\rangle &=& 4 \pi   \left(1 - \frac{1}{R_0} \right) \sqrt{D \, \gamma \,  (R_0-1)} \,t  
\label{mean_perimeter_super} \\
\langle A(t \gg t^*)\rangle &=&  4  \pi   \left(1 - \frac{1}{R_0} \right) D \, \gamma \,  (R_0-1) \,t^2    .
\label{mean_area_super}
\end{eqnarray}
The ballistic growth of the convex hull stems from an underlying travelling front solution of the non-linear equation governing the convex hull behavior. Indeed, the prefactor of the perimeter growth is proportional to the front velocity $v^* = 2 \sqrt{D \, \gamma \,  (R_0-1)}$. As time increases, the susceptible population decreases due to the growth of the infected individuals: this depletion effect leads to a breakdown of the outbreak regime and to a slowing down of the epidemic propagation.

\section{The statistics of the convex hull}

Characterizing the fluctuating geometry of $C$ is a formidable task even in absence of 
branching ($b=0$) and death ($\gamma=0$), i.e., purely for diffusion process
in two dimensions.
Major recent breakthroughs have nonetheless been obtained for 
diffusion processes~\cite{RMC, extremv}
by
a clever adaptation of the Cauchy's integral geometric formulae for
the perimeter and area of any closed convex curve in two dimensions.
In fact,
the problem of computing the mean perimeter and area
of the convex hull of {\em any generic two dimensional} stochastic process
can be mapped, using Cauchy's formulae, to the problem of
computing  
the moments of the maximum and the time at which the maximum occurs 
for the associated one dimensional component stochastic process~\cite{RMC, extremv}.
This was used for computing, e.g., the mean perimeter and area of the
convex hull of a two dimensional regular Brownian motion~\cite{RMC, extremv} and
of a two dimensional random acceleration process~\cite{RAC}.

Our main idea here is to 
extend this method to compute the convex hull statistics for
the two dimensional branching Brownian motion.  
Following this 
general mapping and using isotropy in space (see Supplementary Materials), 
the average perimeter and area of the convex 
hull are
given by
\begin{eqnarray}
\langle L(t)\rangle &=&2\pi \langle x_m(t)\rangle  \label{formula_cauchy_averaged}\\
\langle A(t)\rangle &=& \pi \left[\langle x_m^2(t)\rangle-\langle y^2(t_m)\rangle\right] ,
\label{formula_cauchy_area}
\end{eqnarray}
where $x_m$ is the maximum displacement of our two-dimensional stochastic process
in the $x$ direction up to time $t$,
$t_m$ is the time at which the maximum displacement along $x$ direction occurs
and $y(t_m)$ is the ordinate of the process at $t_m$, i.e., when the
displacement along the $x$ direction is maximal. A schematic representation is provided in Fig.~\ref{fig4}, where the global maximum $x_m$ is achieved by one single infected 
individual, whose path is marked in red. A crucial observation is that the $y$ component of the trajectory connecting $O$ to this red path is a regular one dimensional Brownian motion. Hence, given $t_m$ and $t$,
clearly $\langle y^2(t_m)\rangle=2D\langle t_m\rangle $. 
Therefore,
\begin{equation}
\langle A(t)\rangle = \pi \left[\langle x_m^2(t)\rangle-2D\langle t_m(t)\rangle\right].
\label{avarea}
\end{equation}
Equations~[\ref{formula_cauchy_averaged}] and~[\ref{avarea}] thus show that the 
mean perimeter and area of the epidemics outbreak are related to the extreme 
statistics of a one 
dimensional branching Brownian motion with death.
Indeed, if we can compute the joint distribution $P_t(x_m,t_m)$, we can
in turn compute the three moments $\langle x_m\rangle$, $\langle x_m^2\rangle$ and $\langle 
t_m\rangle$ that are needed in Eqs.~[\ref{formula_cauchy_averaged}] and~[\ref{avarea}].
We show below that this can be performed exactly.

\subsection{The convex hull perimeter and the maximum $x_m$}

For the average perimeter, we just need
the first moment $\langle x_m(t)\rangle=\int_0^{\infty} x_m\, q_t(x_m)\, dx_m$, where
$q_t(x_m)$ denotes  the probability density of the of the maximum of the one dimensional component process.
It is convenient to consider the cumulative distribution $Q_t(x_m)$
i.e., the probability  
that the maximum $x$-displacement 
stays 
below a given value $x_m$ up to time $t$. Then, $q_t(x_m)= dQ_t(x_m)/dx_m$ and
$\langle x_m(t)\rangle= \int_0^{\infty} [1-Q_t(x_m)]\, dx_m$.  Since the process starts at the origin, its maximum
$x$-displacement, for any time $t$, is necessarily nonnegative, i.e., $x_m\ge 0$.
We next write down a backward Fokker-Planck equation describing the
evolution of $Q_t(x_m)$ by considering the three mutually exclusive
stochastic moves in a small time interval $dt$: starting at the origin
at $t=0$,
the walker during the subsequent interval $[0,dt]$ 
dies
with probability $\gamma dt$, infects another individual (i.e., branches) 
with probability $b\, dt=R_0 \gamma dt$, or diffuses by a random displacement $\Delta 
x=\eta_x(0)\, dt$ with 
probability $1-\gamma (1+R_0) dt$. In the last case, its new starting position
is $\Delta x$ for the subsequent evolution.  
Hence, for all $x_m\ge 0$, one can write
\begin{eqnarray}
Q_{t+dt}(x_m)=\gamma dt  + R_0 \gamma dt Q_t^2 (x_m) \nonumber \\
+ [1-\gamma (R_0+1) ]dt \langle Q_t(x_m-\Delta x) \rangle,
\label{Q_feynman}
\end{eqnarray}
where the expectation $\langle \rangle$ is taken with respect to the 
random displacements $\Delta x$.
The first term means that if the process dies right at the start, its maximum up to $t$ is clearly $0$ and
hence is necessarily less than $x_m$. The second term denotes the fact that in case of branching the maximum
of each branch stays below $x_m$: since the branches are independent, one
gets a square. The third term corresponds to diffusion. 
By using $\langle \Delta x \rangle =0 $ and $\langle \Delta x^2 \rangle = 2 D dt$ and 
expanding Eq.~[\ref{Q_feynman}] to the first order in $dt$ and second order in $\Delta x$ 
we obtain
\begin{equation}
\frac{\partial}{\partial t}Q= D \frac{\partial^2}{\partial x_m^2}Q- \gamma (R_0+1) Q + 
\gamma R_0 Q^2 + \gamma 
\label{backward_eq_Q}
\end{equation}
for $x_m\ge 0$, satisfying the boundary conditions $Q_t(0)=0$
and $Q_t(\infty)=1$, and the initial condition 
$Q_0(x_m)=\Theta(x_m)$, where $\Theta$ is the Heaviside step function. 
 Hence from Eq.~[\ref{formula_cauchy_averaged}]
\begin{equation}
\langle L(t)\rangle = 2\pi \int_0^{\infty} [1-Q_t(x_m)]dx_m.
\label{ave_L_def}
\end{equation}
Equation [\ref{backward_eq_Q}] can be solved numerically for all $t$ and all $R_0$, which allows subsequently computing 
$\langle L(t)\rangle$ in Eq.~[\ref{ave_L_def}] (details and figures are provided in the Supplementary Material).

\subsection{The convex hull area}

To compute the average area in Eq.~[\ref{avarea}], we need to evaluate
$\langle x_m^2(t)\rangle$ as well as $\langle t_m\rangle$. Once the cumulative distribution $Q_t(x_m)$ is known,
the second moment $\langle x_m^2(t)\rangle$
can be directly computed by integration, namely, $\langle x_m^2(t)\rangle = \int_0^\infty dx_m 2x_m(1-Q_t(x_m))$.
To determine $\langle t_m\rangle$, we need to also compute the probability density $p_t(t_m)$ of 
the random variable $t_m$.
Unfortunately, writing down a closed equation for $p_t(t_m)$
is hardly feasible. Instead, we first define $P_t(x_m,t_m)$ as the joint probability density
that the maximum of the $x$ component achieves the value $x_m$ at time $t_m$, when
the full process is observed up to time $t$. Then, we derive a backward evolution equation for $P_t(x_m,t_m)$ and then integrate out $x_m$ to derive the marginal density $p_t(t_m)=\int_0^{\infty} P_t(x_m,t_m)\, dx_m$. Following the same arguments as those used for $Q_t(x_m)$ yields a 
backward equation for $P_t(x_m,t_m)$:
\begin{eqnarray}
P_{t+dt}(x_m,t_m)=\left[1-\gamma (R_0+1) dt \right] \langle P_t(x_m-\Delta x, t_m-dt) \rangle \nonumber \\
+ 2 \gamma R_0 dt Q_t(x_m)  P_t(x_m, t_m-dt). 
\label{backward.1}
\end{eqnarray}
The first term at the right hand side represents the contribution from diffusion. The second term 
represents the contribution from branching: we require that one of them attains the maximum 
$x_m$ at the time $t_m - dt$, whereas the other stays below $x_m$ ($Q_t(x_m)$ being the 
probability that this condition is satisfied). The factor $2$ comes from the 
interchangeability of the particles. Developing 
Eq. [\ref{backward.1}] to leading
order gives
\begin{eqnarray}
 \left[\frac{\partial}{\partial t}+\frac{\partial}{\partial t_m}\right]P_t=
\left[D \frac{\partial^2}{\partial x_m^2}- \gamma (R_0+1) + 2\,\gamma\, R_0\, 
Q_t\right]P_t 
\,.
\label{backward_eq}
\end{eqnarray}
This equation describes the time evolution
of $P_t(x_m,t_m)$ in the region $x_m\ge 
0$ and $0\le t_m\le t$. It starts
from the initial condition $P_0(x_m,t_m)=  
\delta(x_m)\,\delta(t_m)$
(since the process begins with a single infected with $x$ component located at $x=0$, it 
implies that at $t=0$ the maximum $x_m=0$ and also $t_m=0$).
For any $t>0$ and $x_m>0$, we have the condition
$P_t(x_m,0)= 0$.
We need to also specify the boundary conditions at $x_m=0$
and $x_m\to \infty$, which read
(i) $P_t(\infty,t_m)=0$ (since
for finite $t$ the maximum is necessarily finite) and 
(ii) $P_t(0,t_m)=\delta(t_m)\,q_t(x_m)\vert_{x_m=0}$. The latter condition comes from the fact 
that,
if $x_m=0$, this corresponds to the event that the $x$ component of the entire process, 
starting at $0$ initially,
stays below $0$ in the time interval $[0,t]$, which happens with probability
$q_t(x_m)\vert_{x_m=0}$: consequently, $t_m$ 
must necessarily be $0$. 
Furthermore, by integrating $P_t(x_m,t_m)$ with respect to $t_m$ we recover the 
marginal density $q_t(x_m)$.

The numerical integration of the full Eq.~[\ref{backward_eq}] would be rather cumbersome. 
Fortunately, we do not need this. 
Since we are only interested in $\langle t_m\rangle $,
it is convenient to introduce
\begin{equation}
T_t(x_m)=\int_0^t t_m P_t(x_m,t_m) dt_m,
\label{eq_ttm}
\end{equation}
from which the average follows as $\langle t_m\rangle=\int dx_m T_t(x_m)$.
Multiplying Eq.~[\ref{backward_eq}] by $t_m$ and integrating by parts we get
\begin{equation}
\frac{\partial}{\partial t}T_t-q_t(x_m)=\left[D \frac{\partial^2}{\partial x_m^2}+2\gamma R_0 
Q_t-\gamma\,(R_0+1)\,\right]T_t,
\label{eq_TTT}
\end{equation}
with the initial condition $T_0(x_m)=0$, and the boundary conditions $T_t(0)=0$ and 
$T_t(\infty)=0$. Eq.~[\ref{eq_TTT}] can be integrated numerically, together with 
Eq.~[\ref{backward_eq_Q}] (details are provided in the Supplementary Material), and the behavior of
\begin{equation}
\langle A(t)\rangle= \pi \int_0^\infty dx_m \left[ 2x_m(1-Q_t(x_m)) - T_t(x_m)\right] 
\label{area_A_def}
\end{equation}
as a function of time is illustrated in Fig.~\ref{fig2}.

\subsection{The critical regime}

We now focus on the critical regime $R_0=1$. We begin with the average perimeter: for $R_0=1$, Eq.~[\ref{backward_eq_Q}]
admits a stationary solution as $t \to \infty$, which can
be obtained by setting $\partial Q/{\partial t}=0$ and solving the resulting differential equation. In fact,
this stationary solution was already known  
in the context of the genetic propagation of a mutant allele~\cite{sawyer}. 
Taking the derivative of this solution with respect to $x_m$, we get the stationary probability density of the maximum $x_m$
\begin{equation}
q_{\infty}(x_m) = \partial_{x_m} Q_{\infty}(x_m)=
\frac{2\sqrt{\frac{\gamma}{6D}}}{\left( 1+\sqrt{\frac{\gamma}{6D}}x_m \right)^3}.
\label{asymptotic_maximum}
\end{equation}
The average is $\langle x_m \rangle = \int_0^{\infty} x_m\, q_{\infty}(x_m)\, dx_m=\sqrt{6D/\gamma}$, which yields then
Eq.~[\ref{mean_perimeter}] for the average perimeter of the convex hull at late times.

To compute the average area in Eq.~[\ref{avarea}], we need to also evaluate
the second moment $\langle x_m^2(t)\rangle$, which diverges as $t\to \infty$,
due to the power-law 
tail of the stationary probability density $q_{\infty}(x_m) \propto x_m^{-3}$ for large 
$x_m$. Hence, we need to consider 
large but finite $t$. In this case, the time dependent probability density
$q_t(x_m)$ displays a scaling form which can be conveniently
written as
\begin{equation}
q_t(x_m) \simeq  q_{\infty}(x_m)  f \! \left( \frac{x_m}{\sqrt{D t}}\right),
\label{scaling_form1}
\end{equation}
where $f(z)$ is a rapidly decaying function with $f(z \ll 1) \simeq 1$, and $f(z \gg 1) 
\simeq 0$. Using the scaling form of Eq.~[\ref{scaling_form1}] and Eq.~[\ref{backward_eq_Q}]
one can 
derive a differential equation for $f(z)$. But it turns out
that we do not really need the solution of $f(z)$.

From Eq.~[\ref{scaling_form1}] we see that the asymptotic power-law decay of $q_t(x_m)$ for 
large $x_m$ has a cut-off
around $x_m^*\sim \sqrt{D t}$ and $f(z)$ is the cut-off function. The
second moment at finite but large times $t$ is given by
$\langle x_m^2(t)\rangle = \int_0^{\infty} x_m^2 q_t(x_m)\, dx_m$.
Substituting the scaling form and cutting off the integral over $x_m$
at $x_m^*=c \sqrt{t}$ (where the constant $c$ depends on the precise form
of $f(z)$) we get, to leading order for 
large $t$,
\begin{equation}
\langle x^2_m(t)\rangle \simeq \int^{x_m^*}_0 x^2_m\, q_{\infty}(x_m)\, dx_m \simeq 
\frac{6D}{\gamma } \ln t\; . 
\label{xm_asymp}
\end{equation}
Thus, interestingly the leading order result is universal, i.e, independent
of the details of the cut-off function $f(z)$ (the $c$-dependence 
is only in the subleading term).
To complete the characterization of $\langle A(t)\rangle $ in Eq.~[\ref{avarea}], we still need to determine $\langle t_m\rangle$:
in the Supplementary Material we explicitly determine the stationary solution
$P_{\infty}(x_m,t_m)$ for $R_0=1$. By following the same arguments as for $\langle x^2_m(t)\rangle$, we show that
\begin{equation}
\langle t_m\rangle \simeq\frac{3}{5 \gamma} 
\ln t 
\label{tm_asymp}
\end{equation}
for large $t$, which leads again to a logarithmic divergence in time. Finally, substituting Eqs.~[\ref{xm_asymp}] and~[\ref{tm_asymp}] in Eq.~[\ref{avarea}]
gives the result announced in Eq.~[\ref{mean_area}].

A deeper understanding of the statistical properties of the process would demand knowing the full distribution
${\rm Prob}(L,t)$ and ${\rm Prob}(A,t)$ of the perimeter and area. These seem rather
hard to compute, but one can obtain the 
asymptotic tails of the distributions by resorting to scaling arguments.
Following the lines of Cauchy's formula (see the Supplementary Material),
it is reasonable to assume that for
each sample the perimeter scales as $L(t)\sim x_m(t)$.
We have seen that the distribution of $x_m(t)$
has a power-law tail for large $t$: $q_{\infty}(x_m)\sim x_m^{-3}$ for
large $x_m$. Then, assuming the scaling $L(t)\sim x_m(t)$ and using 
${\rm Prob}(L,t\to \infty)\, dL\sim
q_{\infty}(x_m)\, dx_m$, it follows that at late times the 
perimeter distribution also has a power-law tail: ${\rm Prob}(L,t\to \infty)\sim L^{-3}$ for 
large $L$.
Similarly, using the Cauchy formula for the area, we can reasonably assume that
for each sample $A(t) \sim x_m^2(t)$ in the scaling regime. Once again, using
${\rm Prob}(A,t\to \infty)\, dA= q_{\infty}(x_m)\, dx_m$, we find that the area
distribution also converges, for large $t$, to a stationary distribution
with a power-law tail: ${\rm Prob}(A,t\to \infty)\sim A^{-2}$ for large $A$. Moreover,
the logarithmic divergence of the mean area calls for a precise ansatz on the tail of the
area distribution, namely,
\begin{equation}
{\rm Prob}(A,t) \xrightarrow[A \gg 1]{} \frac{24 \pi D}{5 \gamma } A^{-2} h \! \left( \frac{A}{D t}\right),
\label{eq_scaling}
\end{equation}
where the scaling function $h(z)$ satisfies the conditions $h(z \ll 1) = 1$, and $h(z \gg 1) 
\simeq 0$. It is not difficult to verify that this is the only scaling compatible with Eq.~[\ref{mean_area}].
These two results are consistent with the fact that for each sample
typically $A(t)\sim L^2(t)$ at late times in the scaling regime. Our scaling predictions are in agreement with our Monte Carlo simulations (see Fig.~\ref{fig2}). The
power-law behavior of ${\rm Prob}(A,t)$ implies that the average area is not representative of
the typical behavior of the epidemic area, which is actually dominated by fluctuations and rare
events, with likelihood given by Eq.~[\ref{eq_scaling}].

\subsection{The supercritical regime}

When $R_0>1$, it is convenient to rewrite Eq.~[\ref{backward_eq_Q}] in terms of $W(x_m,t)=1-Q(x_m,t)$:
\begin{equation}
\frac{\partial}{\partial t} W= D\, \frac{\partial^2}{\partial x_m^2} W + \gamma(R_0-1) W- 
\gamma R_0 W^2
\label{Weq}
\end{equation}
starting from the initial condition $W(x_m,0)=0$ for all $x_m>0$ (see Fig.~\ref{fig3}). 
From Eq.~[\ref{ave_L_def}], $\langle L(t)\rangle = 2\pi \int_0^{\infty} W(x_m,t)\, dx_m$
is just the area under the curve $W(x_m,t)$ vs. $x_m$, up to a factor $2\pi$.
As $t\to \infty$, the system approaches a stationary state for all $R_0\ge 1$, which can be obtained
by setting $\partial_t W=0$ in Eq.~[\ref{Weq}]. For $R_0>1$
the stationary solution $W(x_m,\infty)$ approaches the constant $1-1/R_0$ exponentially fast as
$x_m\to \infty$, namely, $W(x_m,\infty) - 1+R^{-1}_0 \to \exp[-x_m/\xi]$, with a characteristic length scale
$\xi= \sqrt{D/{\gamma(R_0-1)}}$. However, for finite but large $t$, $W(x_m,t)$ 
as a function of $x_m$ has a two-step form: it first decreases from $1$ to its
asymptotic stationary value $1-1/R_0$ over the length scale $\xi$, and then
decreases rather sharply from $1-1/R_0$ to $0$. 
The frontier between the stationary asymptotic value $1-1/R_0$ (stable) and $0$ 
(unstable) moves forward with time at constant velocity, thus creating 
a travelling front at the right end, which separates the stationary value $1-1/R_0$ to the left
of the front and $0$ to the right. 
This front advances with a constant velocity $v^*$ that can be estimated 
using the standard velocity selection principle~\cite{vanSaarloos, DB, MK}. 
Near the front where the nonlinear term is negligible, the equation admits
a travelling front solution: $W(x_m,t)\sim \exp[-\lambda (x_m- v\, t)]$,
with a one parameter family of possible velocities $v(\lambda)= D\lambda + 
\gamma(R_0-1)/\lambda$, parametrized by $\lambda$. This dispersion relation $v(\lambda)$ 
has a minimum at
$\lambda=\lambda^*= \sqrt{\gamma(R_0-1)/D}$, where $v^*= v(\lambda^*)= 2\sqrt{D 
\gamma(R_0-1)}$. According to the standard velocity selection 
principle~\cite{vanSaarloos, DB, MK}, for a sufficiently 
sharp initial condition the system will choose this 
minimum velocity $v^*$. The width of the front remains of $\sim {\cal O}(1)$ at large $t$.
Thus, due to this sharpness of the front, to leading order for large $t$ one can 
approximate $W(x_m,t)\simeq (1-1/R_0)\Theta(v^* t- x_m)$ near the front. Hence, to 
leading order for large $t$ one gets $\langle x_m(t)\rangle \simeq (1-1/R_0) v^* t$ and
$\langle x_m^2 \rangle \simeq (1-1/R_0)\, (v^* t)^2$. 
The former gives, from Eq.~[\ref{formula_cauchy_averaged}], the result announced in Eq.~[\ref{mean_perimeter_super}].
For the mean area in Eq.~[\ref{avarea}], the term $\langle x_m^2 \rangle\sim t^2$ for large $t$ dominates over $\langle t_m \rangle \sim t$ (which can
be neglected), and we get the result announced in Eq.~[\ref{mean_perimeter_super}].
   
\section{Conclusions}

In this paper, we have developed a general procedure for 
assessing the time evolution of the convex hull associated
to the outbreak of an epidemic. We find it extremely appealing that one can successfully use mathematical formulae 
(Cauchy's) from two dimensional integral
geometry to describe the spatial extent of an epidemic outbreak in relatively realistic
situations.
Admittedly, there are many assumptions in this epidemic model 
that are not quite realistic. For instance, we have
ignored the fluctuations of the susceptible populations during the early stages of the epidemic: this hypotheses clearly
breaks down at later times, when depletion effects begin to appear, due to the epidemic invading a
thermodynamical fraction of the total population. In addition, we have assumed that the susceptibles are
homogeneously distributed
in space, which is not the case in reality. Nonetheless, it must be noticed that in practical applications whenever
strong heterogeneities appear, such as mountains, deserts or oceans, one can split the analysis of the evolving
phenomena by conveniently resorting to several distinct convex hulls, one for each separate region.
For analogous reasons, the convex hull approach would not be suitable to characterize
birth-death processes with long range displacements, such as for instance branching L\'evy flights.

The model discussed in this paper based on branching Brownian motion is amenable to exact results. More generally, realistic models
could be taken into account by resorting to cumbersome Monte Carlo simulations. The approach proposed in this paper paves the way
for assessing the spatial dynamics of the epidemic by more conveniently solving two coupled nonlinear equations, under the assumption
that the underlying process be rotationally invariant.

We conclude with an additional remark.
In our computations of the mean perimeter and area, we have averaged over all realizations of the epidemics up to time $t$, including those 
which are already extinct at time $t$. It would also be interesting to consider averages only over the ensemble of epidemics 
that are still active at time $t$. 
In this case we expect different scaling laws for the mean perimeter and the mean area of the convex hull. In particular, in the critical case, we believe that the behavior would be much closer to that of a regular Brownian motion.




\appendix[Supplementary Materials]

\subsection{Cauchy's formula}

The problem of determining the perimeter and the area of the convex hull of any two 
dimensional stochastic process $[x(\tau),y(\tau)]$ with $0\le \tau\le t$ can be mapped to that of computing the 
statistics of the maximum and the time of occurrence of the 
maximum of the one dimensional component process $x(\tau)$~\cite{RMC, extremv}.
This is achieved by resorting to a formula due to Cauchy, which applies to any closed convex 
curve $C$.

A sketch of the method is illustrated in Fig. 5. Choose the coordinates system such that the origin is inside the curve $C$ and 
take a given direction $\theta$. For fixed $\theta$, consider a stick perpendicular to this 
direction and imagine bringing the stick from infinity and stop upon first touching the curve 
$C$. At this point, the distance $M(\theta)$ of the stick from the origin is called the 
support function in the direction $\theta$. Intuitively, the support function measures how 
close can one get to the curve $C$ in the direction $\theta$, coming from infinity. Once the 
support function $M(\theta)$ is known, then Cauchy's formulas~\cite{Cauchy} give the 
perimeter $L$ and the area $A$ enclosed by $C$, namely
\begin{eqnarray}
& L=\int_0^{2\pi} M(\theta)\, d\theta \nonumber \\
& A=\frac{1}{2} \int_0^{2\pi}\left[M^2(\theta)- (M'(\theta))^2\right]d\theta,
\end{eqnarray}
where $M'(\theta)= dM/d\theta$. For example, for a circle of radius $R=r$, $M(\theta)=r$, and one 
recovers the standard formulae: $L=2\pi r$ and $A=\pi r^2$. When $C$ is the convex hull of 
associated with the process at time $t$, we first need to compute its associated 
support function $M(\theta)$. A crucial point is to realize that actually 
$M(\theta)=\max_{0\le \tau \le t} \left[ x(\tau) \cos(\theta)+y(\tau)\sin(\theta) 
\right]$~\cite{RMC, extremv}. Furthermore, if the process is rotationally invariant  any average is 
independent of the angle $\theta$. Hence for the average perimeter we can simply set $\theta=0$ and write $\langle L(t) \rangle=2\pi \langle M(0) \rangle$, where brackets 
denote the ensemble average over realizations. Similarly for the average area,
 $\langle A(t)  \rangle=\pi\left[\langle M^2(0)\rangle- \langle M'(0)^2 \rangle\right]$.
Clearly, $M(0)= \max_{0\le \tau \le t}[x(\tau)]$ is then the maximum of the one dimensional component process $x(\tau)$ for 
$\tau\in[0,t]$. Assuming that $x(\tau)$ takes its maximum value 
$x(t_m)$ at time $\tau=t_m$ (see Fig.~4). Then, $M(0) = x(t_m)= x_m (t) $, and $M'(0) = 
y(t_m) $~\cite{footnote}. Now, by taking the average over Cauchy's formulas, and using 
isotropy, we simply have Eqs. [5] and [6] for the mean perimeter and the mean area 
of the convex hull $C$ at time $t$. Note that this argument is very general and is applicable to
any rotationally invariant two dimensional stochastic process. Since the branching 
Brownian motion with death is rotationally invariant we can use these formulae.

\subsection{Numerical methods}

{\it Numerical integration.} Equations [9] and [14] have been integrated numerically by finite differences in the following way. Time has been discretized by setting $t=n dt$, and space by setting $x=i dx$, where $dt$ and $dx$ are small constants. For the sake of simplicity, here we consider the case $R_0=1$. We thus have
\begin{eqnarray}
Q_{n+1}(i)=\nonumber \\
=Q_n(i)+\gamma \, dt \left[ 1-Q_n(i) \right] ^2 +\nonumber \\
 D\frac{dt}{(dx)^2} \left[ Q_n(i+1) - 2Q_n(i) +Q_n(i-1) \right]
\end{eqnarray}
and
\begin{eqnarray}
T_{n+1}(i)=\nonumber \\
=T_n(i)+2\, \gamma \, dt \, T_n(i) \left[ Q_n(i)-1 \right] + \nonumber \\
 D\frac{dt}{(dx)^2} \left[ T_n(i+1) - 2 T_n(i) +T_n(i-1) \right] +\nonumber \\
 \frac{dt}{dx} \left[ T_n(i)-T_n(i-1) \right].
\end{eqnarray}
As for the initial conditions, $Q_0(0)=0$ and $Q_0(i>0)=1$, and $T_0(i)=0$ $\forall i$. The boundary conditions at the origin are $Q_n(0)=0$ and $T_n(0)=0$. In order to implement the boundary condition at infinity, we impose $Q_n(i_{\max})=1$ and $T_n(i_{\max})=0$ $\forall n$, where the large value $i_{\max}$ is chosen so that $T_n(i_{\max})-T_n(i_{\max}-1)< 10^{-7} $. We have verified that numerical results do not change when passing to the tighter condition $T_n(i_{\max})-T_n(i_{\max}-1)< 10^{-9} $.

Once $Q_n(i)$ and $T_n(i)$ are known, we use Eqs. [10] and [15] to determine the average perimeter and area, respectively.

{\it Monte Carlo simulations.} The results of numerical integrations have been confirmed by running extensive Monte Carlo simulations. Branching Brownian motion with death has been simulated by discretizing time with a small $dt$: in each interval $dt$, with probability $b dt$ the walker branches and the current walker coordinates are copied to create a new initial point, which is then stored for being simulated in the next $dt$; with probability $\gamma dt$ the walker dies and is removed; with probability $1-(b+\gamma)dt$ the walker diffuses: the $x$ and $y$ displacements are sampled from Gaussian densities of zero mean and standard deviation $\sqrt{2 D dt}$ and the particle position is updated. The positions of all the random walkers are recorded as a function of time and the corresponding convex hull is then computed by resorting to the algorithm proposed in \cite{berg}.

{\it Perimeter statistics.} In order the complete the analysis of the convex hull statistics, in Fig~\ref{fig2_SM} and Fig~\ref{fig3_SM} we show the results for the perimeter. 

\subsection{Analysis of $t_m$}

In the critical case $R_0=1$, the stationary joint probability density $P_{\infty}(x_m,t_m)$ satisfies (upon setting $\partial P_t/\partial t=0$ in Eq. [12])
\begin{eqnarray}
 \frac{\partial}{\partial t_m}P_{\infty}(x_m,t_m)=\nonumber \\
 = \left[ D \frac{\partial^2}{\partial x_m^2} 
- \frac{2 \gamma }{\left[1+\sqrt{\frac{\gamma }{6D}} x_m \right]^2} 
\right]P_{\infty}(x_m,t_m)\; .
\label{backward_eq_2}
\end{eqnarray}
For any $x_m>0$, we have the condition $P_{\infty} (x_m,0)=0$. The boundary conditions for Eq. (\ref{backward_eq_2}) are $P_{\infty}(x_m \to \infty, t_m)=0$ and $P_{\infty}(0, t_m) = q_{\infty}(0)\, \delta(t_m) 
= 2 
\sqrt{\gamma/(6D)}\, \delta(t_m)$. We first take the Laplace transform of (\ref{backward_eq_2}), namely,
\begin{equation}
\tilde{P}_{\infty}(x_m,s)=\int_0^\infty e^{-st_m 
}\, P_{\infty}(x_m,t_m)\,dt_m.
\label{laplace.0}
\end{equation}
This gives for
all $x_m>0$
\begin{equation}
 \frac{D}{s}\frac{\partial^2}{\partial x_m^2}\tilde{P}_{\infty}(x_m,s) =\left[ 1 + 
\frac{12}{\frac{s}{D}( \sqrt{\frac{6D}{\gamma}} + x_m )^2} \right] \tilde{P}_{\infty}(x_m,s) ,
\end{equation}
where we have used the condition $P_{\infty}(x_m,0)=0$ for any $x_m>0$.
This second order differential equation satisfies two boundary conditions:
$\tilde{P}_{\infty}(\infty,s)=0$ and 
$\tilde{P}_{\infty}(0,s)=2\sqrt{\gamma/(6D)}$. 
The latter condition is obtained by Laplace transforming $P_{\infty}(0,t_m)= 
2\sqrt{\gamma/(6D)}\, \delta(t_m)$.
By setting
\begin{equation}
z=\left( \sqrt{\frac{6D}{\gamma}} +x_m\right)  \sqrt{\frac{s}{D}},
\end{equation}
we rewrite the equation as
\begin{equation}
 \frac{\partial^2}{\partial z^2}\tilde{P}_{\infty} - \tilde{P}_{\infty} - 
\frac{12}{z^2} \tilde{P}_{\infty} = 0.
\end{equation}
Upon making the transformation $\tilde{P}_{\infty}(z)=\sqrt{z}\,F(z)$, the function 
$F(z)$ then
satisfies the 
Bessel differential equation
\begin{equation}
\frac{d^2}{d z^2}F(z)+\frac{1}{z}\frac{d}{d z}F(z) -\left[ 1+\frac{49}{4 z^2} \right] F(z) =0 .
\end{equation}
The general solution of this differential equation can be expressed as a linear combination of two 
independent solutions: $F(z)= A\, I_{7/2}(z) + B\, K_{7/2}(z)$  
where $I_{\nu}(z)$ and $K_{\nu}(z)$ are modified Bessel functions. Since, $I_{\nu}(z)\sim e^{z}$ for large $z$, 
it is clear that to satisfy the boundary condition $\tilde{P}_{\infty}(\infty,s)=0$ (which mean $F(z\to \infty)=0$), we need to choose $A=0$.
Hence we are left with 
$F(z)=B K_{7/2}(z)$, 
where the constant $B$ is 
determined from the second boundary condition $\tilde{P}_{\infty}(0,s)= 2\sqrt{\gamma/(6D)}$. By reverting to the variable $x_m$, we 
finally get
\begin{equation}
\tilde{P}_{\infty}(x_m,s) = 2 \sqrt{\frac{\gamma}{6D}}\, 
\sqrt{1+\frac{\gamma}{6D}\,x_m}\frac{K_{7/2}\left[ \left( \sqrt{\frac{6D}{\gamma}} 
+x_m\right)  
\sqrt{\frac{s}{D}}\right]}{K_{7/2}\left[ \sqrt{\frac{6s}{\gamma}}  \right]}.
\label{laplace.1}
\end{equation}
Now, we are interested in determining the Laplace transform of the 
marginal density $\tilde{p}_{\infty}(s)=\int_0^\infty e^{-s\,t_m }\, 
p_{\infty}(t_m)\,dt_m$ where $p_{\infty}(t_m)= \int_0^{\infty} P_{\infty}(x_m,t_m)\, dx_m$.
Taking Laplace transform of this last relation with respect to $t_m$ gives
$\tilde{p}_{\infty}(s)=  
\int_0^\infty \tilde{P}_{\infty}(x_m,s)\, dx_m $.
Once we know $\tilde{p}_{\infty}(s)$, we can invert it to obtain $p_{\infty}(t_m)$.
Since we are interested only in the asymptotic tail of $p_{\infty}(t_m)$, it suffices
to investigate the small $s$ behavior of $\tilde{p}_{\infty}(s)$. Integrating Eq. (\ref{laplace.1}) over $x_m$ and
taking the $s\to 0$ limit, we obtain after some straightforward algebra 
\begin{equation}
\tilde{p}_{\infty}(s) = 1 + \frac{3}{5 \gamma}\, s\, \ln (s)+\cdots.
\end{equation}
We further note that 
\begin{equation}
\int_0^\infty e^{-s t_m}\, t_m^2\, p_{\infty}(t_m)\, dt_m = \frac{d^2}{d 
s^2}\tilde{p}_{\infty}(s) \simeq 
\frac{3}{5 \gamma s},
\end{equation}
which can then be inverted to give the following asymptotic behavior for large $t_m$
\begin{equation}
p_{\infty}(t_m) \simeq \frac{3}{5 \gamma t_m^2}\; .
\label{ptm.1}
\end{equation}
Analogously as for $\langle x_m^2 \rangle $, 
the moment $\langle t_m \rangle \to \infty$, due to the power-law 
tail $p_{\infty}(t_m) \propto t_m^{-2}$. 
Hence we need to compute $\langle t_m\rangle$ for large but finite $t$: in this case, the time-dependent solution
displays a scaling behavior
\begin{equation}
p_t(t_m) \simeq  p_{\infty}(t_m)\, g \! \left (\frac{t_m}{t} \right),
\end{equation}
where the scaling function $g(z)$ satisfies the conditions $g(z \ll 1) \simeq 1$ and $g(z \gg 
1) =0$. Much like in Eq.~[17] for the marginal density $q_t(x_m)$,
we have a power-law tail of $p_t(t_m)$ for large $t_m$ that has a cut-off
at a scale $t_m^*\sim t$, and $g(z)$ is the cut-off function. As in the case
of $x_m$, we do not need the precise form of $g(z)$ to compute the leading
term of the first moment $\langle t_m\rangle= \int_0^{\infty} p_t(t_m)\, t_m\, dt_m$ for 
large $t$. Cutting off the integral at $t_m^*= c_1 t$ (where $c_1$ depends on the
precise form of $g(z)$) and performing the integration gives
 \begin{equation}
\langle t_m\rangle \simeq \int^{t}_0 t_m\, p_{\infty}(t_m)\, dt_m \simeq\frac{3}{5 \gamma} 
\ln t ,
\end{equation}
which is precisely the result announced in Eq.~[19].

\begin{acknowledgments}

S.N.M. acknowledges
support from the ANR grant 2011-BS04-013-01 WALKMAT. S.N.M and A.R. 
acknowledge
support from the Indo-French
Centre for the Promotion of Advanced Research under Project 4604-3.

\end{acknowledgments}


\begin{thebibliography}{27}
\bibitem{footnote} Actually, $t_m$ implicitly depends on $\theta$, hence formally $M'(\theta)=- 
x(t_m) \sin(\theta)+ y(t_m)\cos (\theta) + 
\frac{dt_m}{d\theta}\,\frac{dz_{\theta}(t)}{dt}\Big|_{t=t_m}$. Nonetheless, since $z_{\theta}(t)$ is maximum at $t=t_m$, by definition $dz_{\theta}(t)/dt|_{t=t_m}=0$.
\end{thebibliography}




\end{article}









\begin{figure*}
\includegraphics[width=\textwidth]{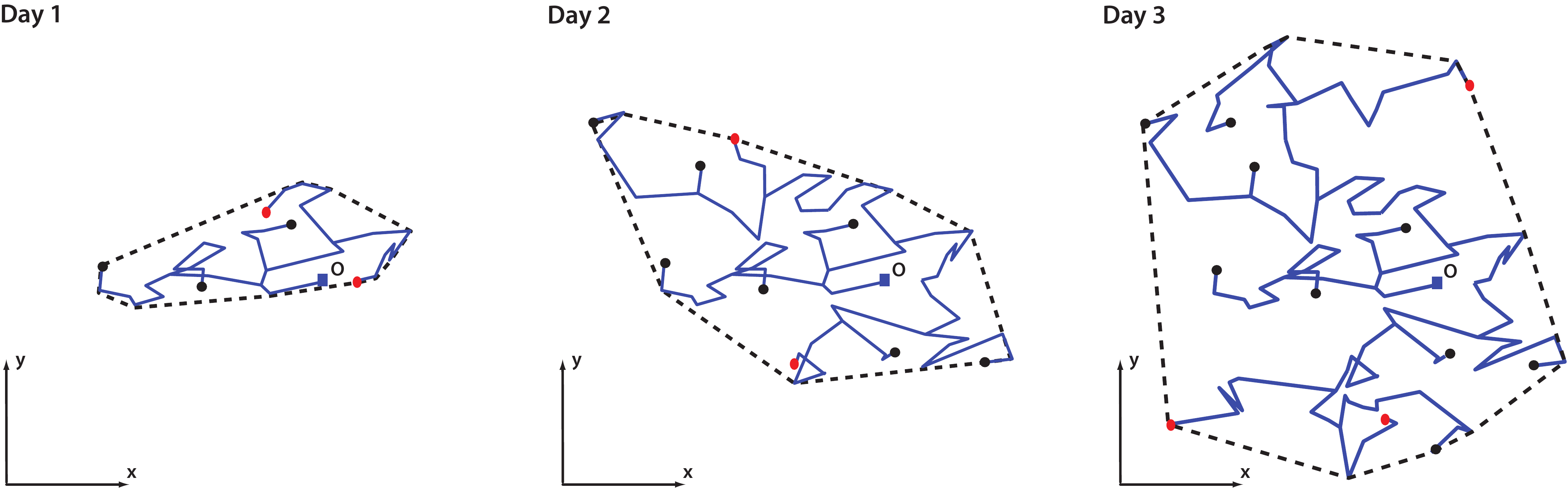}
\caption{The snapshots of the trajectories of an assembly of infected individuals at the epidemics outbreak at three different times (schematic), starting from a single infected at the origin $O$ at time $t=0$. Individuals that are still infected at a given time $t$ are displayed as red dots, while those already recovered are shown as black dots. The convex hull enclosing the trajectories (shown as a dashed line) is a measure of geographical area
covered by the spreading epidemic. As the epidemic grows in space, the associated convex hull also grows in time.}
\label{fig1}
\end{figure*}

\begin{figure*}
\includegraphics[width=\textwidth]{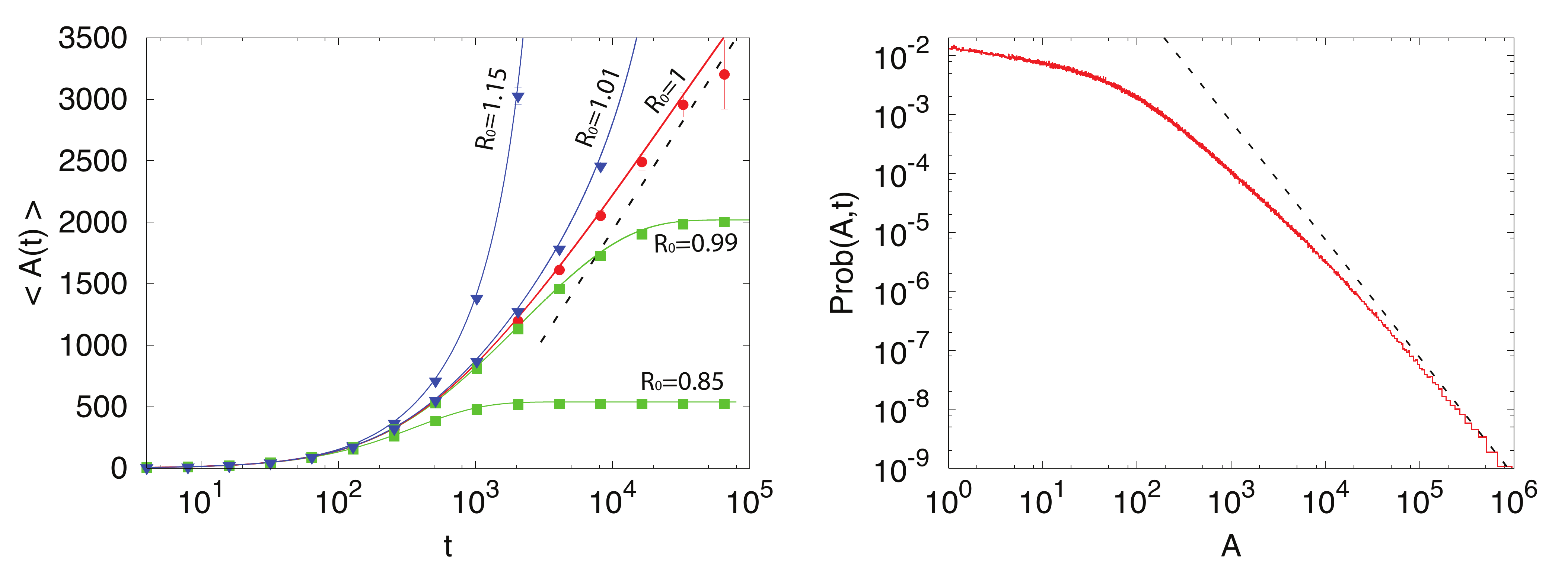}
\caption{Left. The average area $\langle A(t)\rangle $ of the convex hull as a function of the observation time. For the parameter values, we have chosen $D=1/2$ and $b=R_0\gamma=0.01$. We considered five different values of $R_0$. 
We have obtained these results by two different methods: (i) via the numerical integration of Eqs. [\ref{backward_eq_Q}]  and [\ref{eq_TTT}] and using Eq.  [\ref{area_A_def}]. These results are displayed as solid lines. (ii) by Monte Carlo simulations of the
two-dimensional branching Brownian motion with death with the same parameters, averaged over $10^5$ samples. Monte Carlo are displayed as symbols.
The dashed lines represent the asymptotic limits as
given in Eq. [\ref{mean_area}] for the critical case $R_0=1$. Further details of the numerical simulations are provided in the Supplementary Material. Right. Distribution of  the area of the convex hull for the critical case $R_0=1$, with $\gamma=0.01$ and $D=1/2$, as obtained by Monte Carlo simulations with $2\cdot 10^6 $ realizations. The dashed line corresponds to the power-law $(24 \pi D/ 5 \gamma) A^{-2}$ as predicted by Eq [\ref{eq_scaling}].
}
\label{fig2}
\end{figure*}

\begin{figure*}
\includegraphics[width=\textwidth]{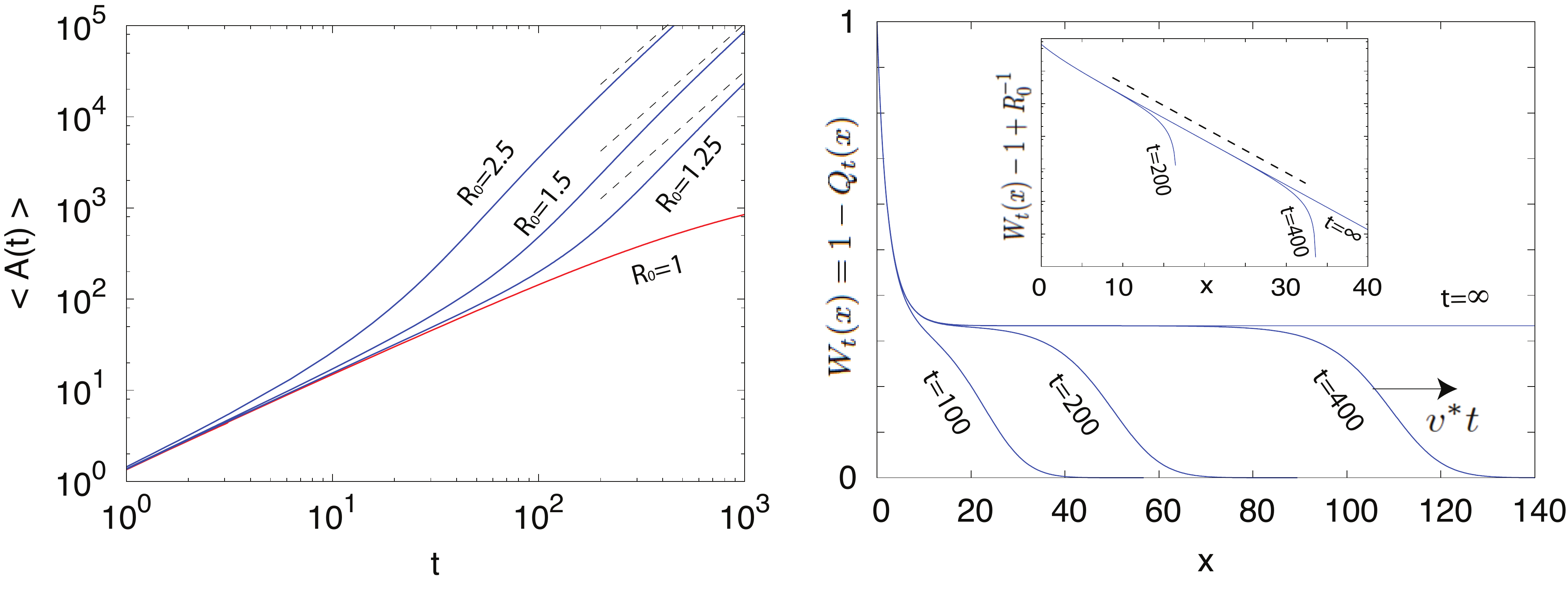}
\caption{Left. The time behavior of the average area in the supercritical regime
for different values of $R_{0}>1$. Dashed lines represent the asymptotic
scaling as in Eq. [\ref{mean_area_super}].
The red curve corresponds to the critical regime. Right. The behavior
of $W_{t}(x)=1-Q_{t}(x)$ for $R_{0}=1.5$ at different times, as in Eq. [\ref{Weq}]. When
$t\rightarrow\infty$, $W_{t}(x)\rightarrow1-R_{0}^{-1}$ , and for
large but finite times the travelling front behavior is clearly visible.
The inset displays the exponential convergence of $W_{t}(x)$ to the
asymptotic limit. The dashed line represents $\xi=\sqrt{D/\gamma(R_{0}-1)}$.
}
\label{fig3}
\end{figure*}

\begin{figure*}
\includegraphics[width=\textwidth]{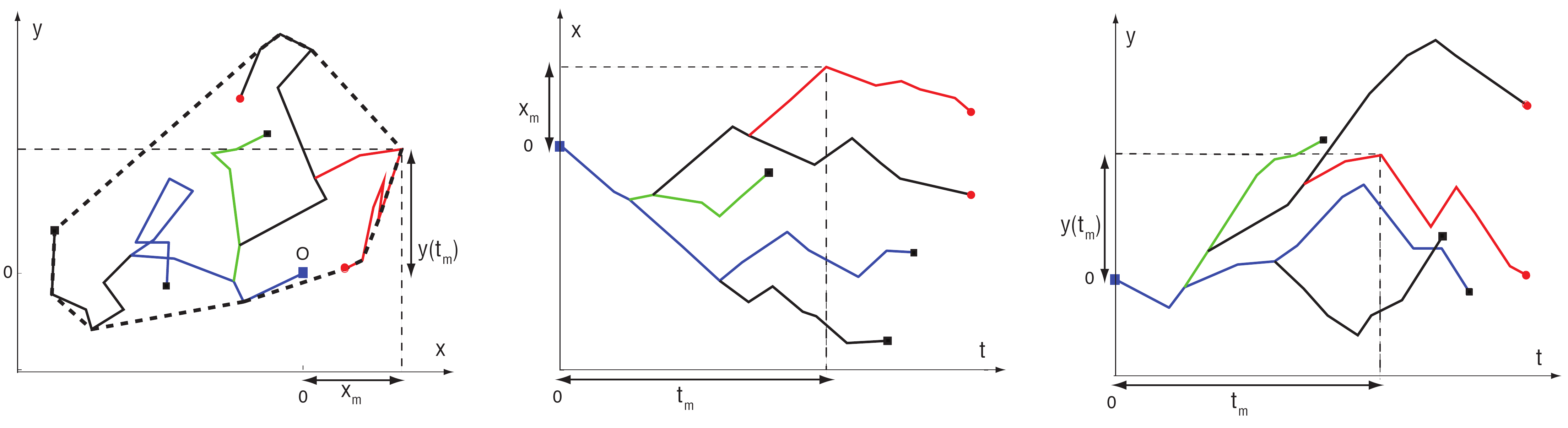}
\caption{Left. A branching random walk composed of five individuals. At time $t=0$, a single infected is at the origin $O$, and starts diffusing (blue line). At later times, this individual branches and gives rise to other infected individuals. Among these, the red path reaches the maximum $x_m$ along the $x$ component up to the final time $t$. Infected individuals at a given time $t$ are displayed as red dots, whereas recovered as black dots.
Center. The displacement along the $x$ direction as a function of time. The red path reaches the global maximum $x_m$ at time $t_m$.
Right. The displacement along the $y$ direction as a function of time. When the red path reaches the global maximum $x_m$ at time $t_m$, its $y$ coordinate attains the value $y(t_m)$. A crucial observation is that the $y$ component of the trajectory connecting $O$ to the red path is a regular Brownian motion. This is not the case for the $x$ component, which is constrained to reach the global maximum of the branching process.}
\label{fig4}
\end{figure*}

\begin{figure}[b!]
\begin{center}
\includegraphics[width=0.6\textwidth]{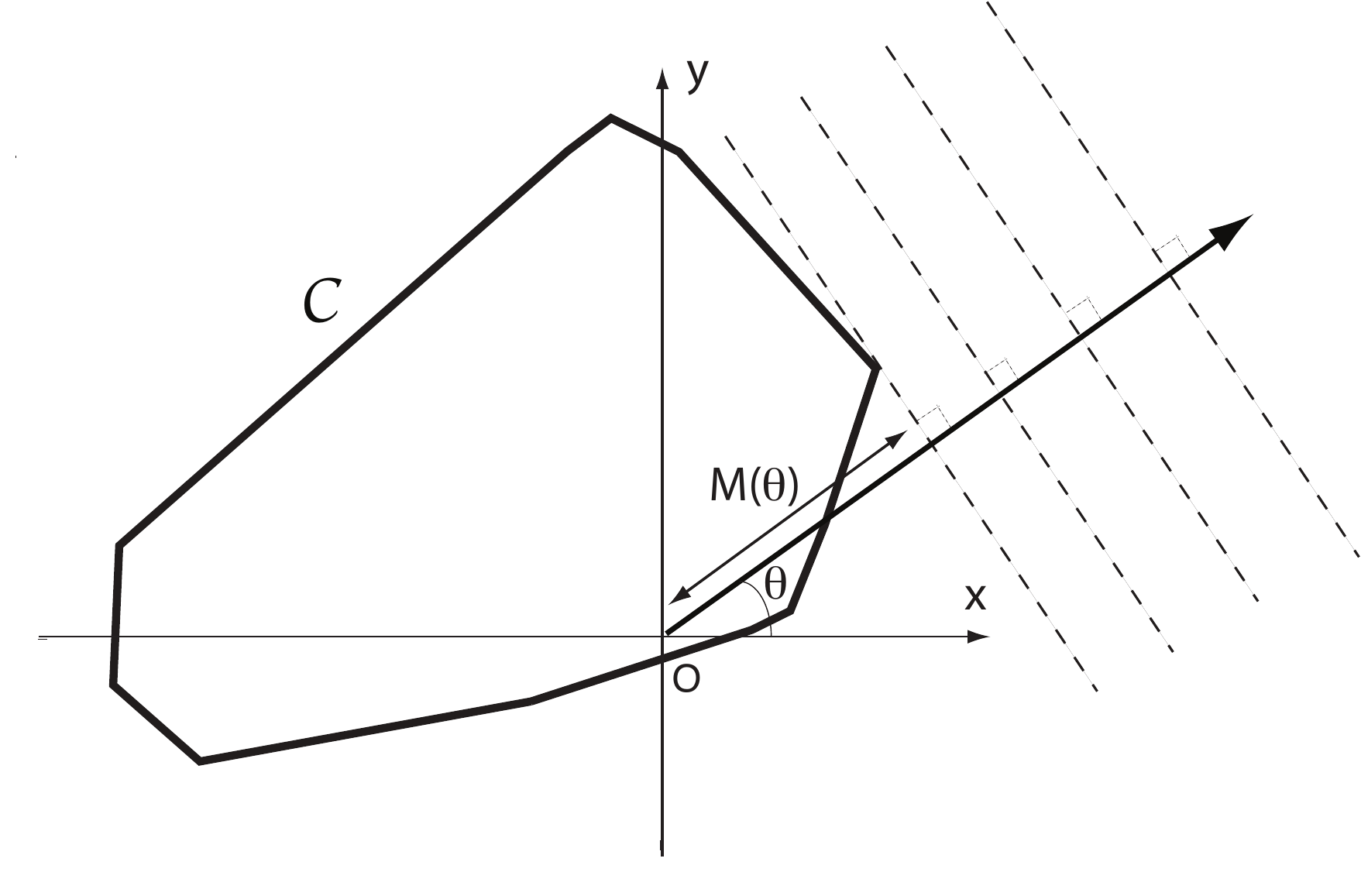}
\caption{Cauchy's construction of the two-dimensional convex hull, with support function $M(\theta)$ representing the distance along the direction $\theta$.}
\label{fig3}
\end{center}
\end{figure}

\begin{figure*}
\includegraphics[width=\textwidth]{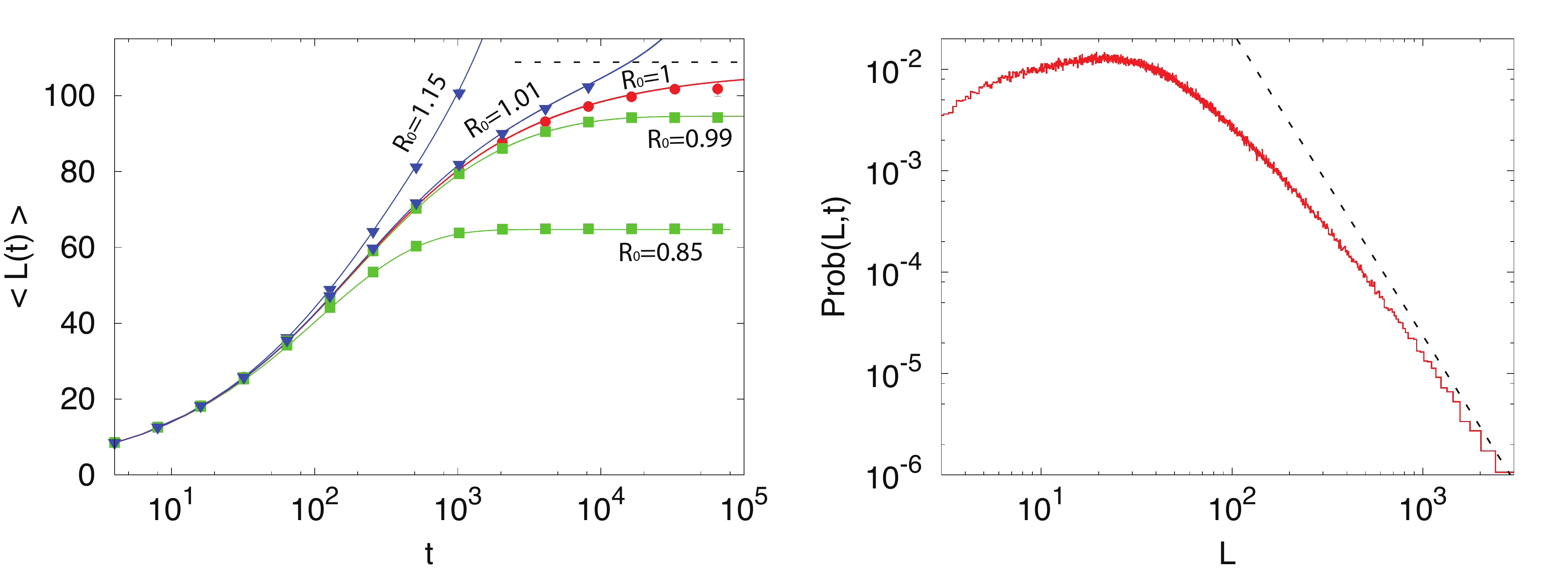}
\caption{Left. The average perimeter $\langle L(t)\rangle $  of the convex hull as a function of 
the observation time. For the parameter values, we have chosen $D=1/2$ and $b=R_0\gamma=0.01$. We considered five different values of $R_0$. We have obtained these results by two different methods: 
(i) via the numerical integration of Eq. [9] and using Eq. [10] (with the choices
$dt=0.003125$ 
and $dx=0.1768$). These results are displayed as solid lines. (ii) by Monte Carlo simulations of the
two-dimensional branching Brownian motion with death with the same parameters and with the choice of the Monte Carlo time step $dt=0.25$
with the results averaged over $10^5$ samples. Monte Carlo are displayed as symbols.
The dashed lines represent the asymptotic limits as given in Eq. [1] for the critical case $R_0=1$. Right.
Distribution of the perimeter  of the convex hull for the critical case $R_0=1$, with $\gamma=0.01$ and $D=1/2$, as obtained by Monte Carlo simulations with
time step $dt=1$ and $t=4\cdot 10^5$. The number of realizations is $2\cdot 10^6 $. The dashed line of the left panel corresponds to the power-law $L^{-3}$ (up to an arbitrary prefactor).}
\label{fig2_SM}
\end{figure*}

\begin{figure*}
\begin{center}
\includegraphics[width=0.5\textwidth]{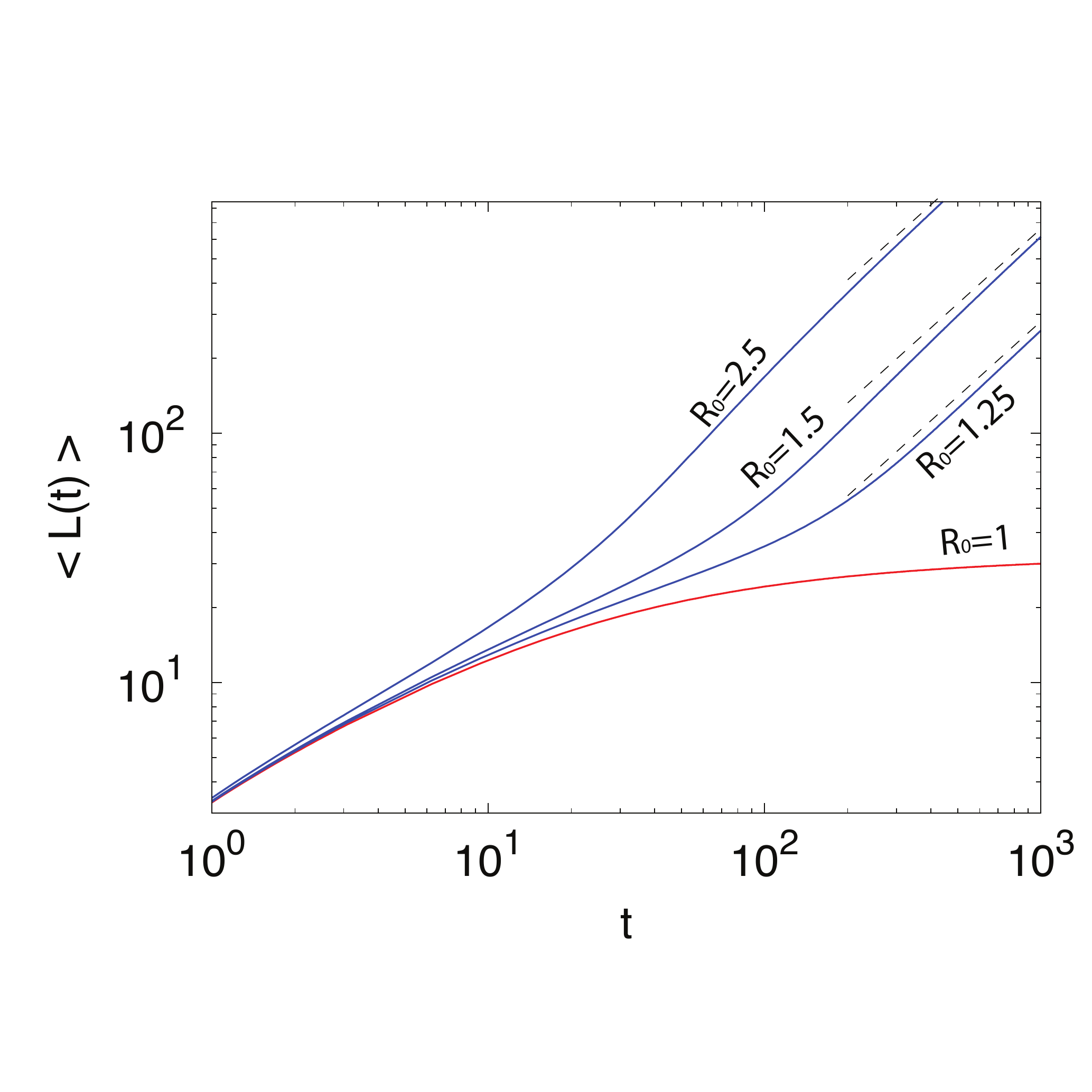}
\caption{The time behavior of the average perimeter in the supercritical regime
for different values of $R_{0}>1$. Dashed lines represent the asymptotic
scaling as in Eq. [3].
The red curve corresponds to the critical regime.
}
\label{fig3_SM}
\end{center}
\end{figure*}

\end{document}